\documentclass[10pt,journal,compsoc]{IEEEtran}
\ifCLASSOPTIONcompsoc
  \usepackage[nocompress]{cite}
\else
  \usepackage{cite}
\fi
\usepackage{multirow}
\usepackage{enumitem}
\usepackage[flushleft]{threeparttable}
\usepackage{tabularx}
\ifCLASSINFOpdf
\usepackage[pdftex]{graphicx}
\graphicspath{{./figures/}}
\DeclareGraphicsExtensions{.pdf,.jpg,.png}
\else
\fi
\usepackage[cmex10]{amsmath}
\newcolumntype{P}[1]{>{\centering\arraybackslash}p{#1}}
\newcolumntype{Y}{>{\centering\arraybackslash}X}
\begin{document}
\title{Self-Adaptive Trade-off Decision Making for Autoscaling Cloud-Based Services}

\author{Tao~Chen,~\IEEEmembership{Member,~IEEE,}
        Rami~Bahsoon,~\IEEEmembership{Member,~IEEE}
        %and~Jane~Doe,~\IEEEmembership{Life~Fellow,~IEEE}% <-this % stops a space
\IEEEcompsocitemizethanks{\IEEEcompsocthanksitem T. Chen and R. Bahsoon are with CERCIA, the School
of Computer Science, University of Birmingham, Birmingham,
UK, B15 2TT.\protect\\
E-mail: t.chen@cs.bham.ac.uk, r.bahsoon@cs.bham.ac.uk
}%\IEEEcompsocthanksitem J. Doe and J. Doe are with Anonymous University.}% <-this % stops an unwanted space
\thanks{Manuscript received 8 Apr. 2015; revised 20 Sept. 2015; accepted 30 Oct.2015. Date of publication 0 . 0000; date of current version 0 . 0000.For information on obtaining reprints of this article, please send e-mail to:reprints@ieee.org, and reference the Digital Object Identifier below.Digital Object Identifier no. 10.1109/TSC.2015.2499770 }}


\markboth{IEEE TRANSACTIONS ON SERVICES COMPUTING,  VOL. 8,  NO. X,  XXXXX 2015}%
{Shell \MakeLowercase{\textit{et al.}}: Bare Demo of IEEEtran.cls for Computer Society Journals}

%\IEEEpubid{\makebox[\columnwidth]{\hfill 0000--0000/00/\$00.00~\copyright~2014 IEEE}%


\IEEEtitleabstractindextext{%
\begin{abstract}
Elasticity in the cloud is often achieved by on-demand autoscaling. In such context, the goal is to optimize the Quality of Service (QoS) and cost objectives for the cloud-based services. However, the difficulty lies in the facts that these objectives, e.g., throughput and cost, can be naturally conflicted; and the QoS of cloud-based services often interfere due to the shared infrastructure in cloud. Consequently, dynamic and effective trade-off decision making of autoscaling in the cloud is necessary, yet challenging. In particular, it is even harder to achieve well-compromised trade-offs, where the decision largely improves the majority of the objectives; while causing relatively small degradations to others. In this paper, we present a self-adaptive decision making approach for autoscaling in the cloud. It is capable to adaptively produce autoscaling decisions that lead to well-compromised trade-offs without heavy human intervention. We leverage on ant colony inspired multi-objective optimization for searching and optimizing the trade-offs decisions, the result is then filtered by compromise-dominance, a mechanism that extracts the decisions with balanced improvements in the trade-offs. We experimentally compare our approach to four state-of-the-arts autoscaling approaches: rule, heuristic, randomized and multi-objective genetic algorithm based solutions. The results reveal the effectiveness of our approach over the others, including better quality of trade-offs and significantly smaller violation of the requirements. 
\end{abstract}

}

\maketitle

\IEEEdisplaynontitleabstractindextext

% \fi
\IEEEpeerreviewmaketitle

\IEEEraisesectionheading{\section{Introduction}\label{sec:introduction}}
\IEEEPARstart{C}{loud} computing, grounded on the principle of shared infrastructure, emerges as an increasingly important computing paradigm. In such paradigm, Software as-a-Service (SaaS) are typically supported by the software stack in the Platform as-a-Service (PaaS) layer. They are also supported with Virtual Machines (VM) and hardware within the Infrastructure as-a-Service (IaaS) layer. When running cloud-based services under changing environmental conditions (e.g., workload, size of incoming job etc.), governing their Quality of Service (QoS) is among the primary concerns of both cloud providers and service owners. The QoS, for examples, can be response time, throughput or any other non-functional attributes experienced by the end-users. In the cloud, these QoS attributes can be often managed and tuned through various internal control knobs, including software configurations (e.g., number of service threads) and hardware resources (e.g., CPU and memory of VM), in a shared infrastructure subject to rental cost. During such process, it is particularly important to consider the interplay between software configurations and hardware resources. This is because these software configurations can significantly affect the QoS and the required resources, as evident by many recent work \cite{2013-JRAO-most-closest-work-2013}\cite{software-RP-two-loops}\cite{2014-decision-tree-software-CP-2014}. 

To achieve elasticity and scalability in cloud, autoscaling is usually considered as the key strategy. In general term, autoscaling refers to the elastic process(es) that adapts the control knobs on-demand according to the changing environment conditions. Its ultimate goal is to continually optimize the QoS and cost objectives for all cloud-based services; thus their requirements can be better complied. Here, the core phase in autoscaling is the dynamic decision making process that produces the optimal (or near-optimal) \textbf{\emph{decision}}\textemdash a set of newly configured values of the necessary control knobs\textemdash for all the related objectives. However, \textbf{\emph{objective-dependency}} (i.e., conflicted or harmonic objectives) often exist in the decisions making process, which implies that trade-offs are necessary and hence it renders the reasoning about the effects of decisions on objectives as a complex task. This is especially true for the shared infrastructure of cloud where objective-dependency exists for both intra- and inter-services. That is to say, trade-off is not only caused by the nature of objectives (intra-service), e.g, Throughput and cost objective of a service; but also by the \textbf{\textit{dynamic QoS interference}}  (inter-services) due to the co-located services on a VM and co-hosted VMs on a Physical Machine (PM) \cite{2013-JRAO-most-closest-work-2013}\cite{software-RP-two-loops}\cite{2014-decision-tree-software-CP-2014}\cite{qcloud}. Here, QoS interference refers to scenarios where a service exhibits wide disparity in its QoS performance that depends on the dynamic behaviors of its neighbors; this is known as a typical consequence of resources contention in cloud \cite{2014-decision-tree-software-CP-2014}\cite{qcloud}. Therefore, given the presence of complex objective-dependency, it is clear that the decision making for autoscaling in the cloud is very difficult, if not impossible, to be handled by human decision makers; and thus urges the need for self-adaptivity. Among the trade-off decisions that quantified by the commonly used pareto-dominance relation, we are particularly interested in the ones that achieve \textbf{\textit{well-compromised trade-offs}} (a.k.a. knee points). A decision is said to result in well-compromised trade-off, as when compared with its neighboring decisions, if it largely improves the majority of the objectives; while causing relatively small degradations to others. In other words, the improvements of all dependent objectives are well-balanced.

The QoS performance of services and the cloud environment tend to fluctuate; consequently, the QoS interference, the possible trade-off decisions for autoscaling and their effects on the objectives are dynamic and uncertain. State-of-the-art approaches often ignore QoS interference and its related trade-offs in autoscaling. Furthermore, they tend to be limited in handling two challenges related to the trade-offs: Firstly, most of the work restricts the autoscaling decisions into fixed bundles (e.g., VM instance), which is rather inflexible, and thus it is necessary to consider any combinations of the configured values for control knobs \cite{parallel-RL-vertical-QoS-2013}. However, given the potentially large amount of possible combinations of the configured values, finding the optimal decisions and reasoning about their effects on objectives is known to be an NP-hard problem \cite{Antonescu13dynamicsla}\cite{E3-R-extended}\cite{GA-full-simulation}. Henceforth, the key challenge is how to dynamically optimize diversified trade-off decisions and thus produce better coverage of the trade-offs surface. Secondly, another challenge is how to dynamically extract the decisions that achieve well-compromised trade-offs under runtime uncertainty. This challenge is not well-studied in cloud autoscaling.

Existing work for autoscaling decision making in the cloud can be either \emph{static} in the sense that the mapping between conditions and decisions are fixed; or \emph{dynamic} where the runtime conditions and behaviors are used to 'learn' new decisions. The static approaches \cite{Compsac_2010_I_Brandic}\cite{scale-rule-based} are insufficient as they are restricted by the simplified assumptions about the conditions and the mapped decisions. Although dynamic approaches have been proposed to address this limitation, most of them \cite{multitier-resalloc-Cloud11}\cite{2014-2-SVM-workload-type-2014}\cite{ILP-cost-only-scaling-2013} only focus on optimizing a single objective (e.g., cost), where other objectives are treated as constraints. This means that the search process tends to be limited in exploring trade-offs due to the optimization of single objective. To this end, weighted-sum formulation that aggregates all the objectives into a single one has been widely applied \cite{cache-static-ANN-bi-obj-2012}\cite{2014-profiling-decision-tree-scaling-2014}\cite{software-RP-two-loops}\cite{smart2004}\cite{Antonescu13dynamicsla}. Nevertheless, weighted-sum of objectives requires human intervention to carefully design and tune the weights for the objectives, which is often an extremely complex and error-prone exercise. In addition, finding the right weights in advance is extremely difficult in the presence of QoS interference, as it is difficult to presume the relative importance of the services and their levels of importance. On the other hand, a single aggregation can track the search in a smaller search space and the resulted decisions are driven by coarser and less information about the trade-offs surface. In other words, the optimality and diversity of the resulted trade-offs decisions tend to be limited and therefore causing it difficult to achieve well-compromised trade-offs. There is a limited amount of work that leverage on the notion of multi-objective optimization \cite{E3-R-extended}\cite{GA-full-simulation}\cite{2014-eplison-GA-weigh-h-scaling-2014} and pareto-dominance \cite{Goldberg} based sort. Most commonly, they apply Multi-Objective Genetic Algorithm (MOGA), e.g., NSGA-II \cite{nsgaii}, to search the trade-offs decisions without explicitly using weights. However, since they do not focus on decisions that produces well-compromised trade-off, the amount of resulted decisions is unavoidably large and can easily lead to imbalanced improvement.

In this paper, we propose a multi-objective self-adaptive approach for autoscaling decision making in the cloud without heavy human intervention. This approach  dynamically and adaptively adjust its own behaviors to (i) optimize and discover the diversified trade-offs decisions at runtime; and (ii) extract the decisions that produce well-compromised trade-offs with respect to all related objectives. To the best of our knowledge, we are the first to address the problem of reaching well-compromised trade-offs for autoscaling in the cloud while considering the trade-offs caused by QoS interference. In particular, we show the effectiveness of the approach for up to 30 dependent objectives, which is significantly larger than what is considered in state-of-the-art work. Precisely, we make the following novel contributions:

\textbf{\textit{Firstly}}, by leveraging on our prior work \cite{Chen:2013}\cite{Chen:2014:ucc}\cite{Chen:2014}, we implement the proposed approach as a self-adaptive and standalone system that is QoS interference aware.

\textbf{\textit{Secondly}}, in light of many successful applications of Metaheuristic Algorithms (MAs) in the cloud \cite{software-RP-two-loops}\cite{Antonescu13dynamicsla}\cite{E3-R-extended}\cite{GA-full-simulation}\cite{ant-cloud}\cite{ant-cloud-vm}, we design Multi-Objective Ant Colony Optimization (MOACO) to search the optimal (or near-optimal) trade-offs decisions for cloud autoscaling. The search process in our MOACO is similar to conduct many single objective optimizations in one run, which aims to optimize and to make trade-off for large number of objectives without specifying weights on them. 

We have chosen MAs over the deterministic algorithms because (i) since the problem tends to be NP-hard, existing deterministic algorithms can perform poorly on high dimensionality as they aim for exact result. On the other hand, the MAs are capable to efficiently achieve approximated results with good enough quality. (ii) Deterministic algorithms often rely on and take advantages from the nature of the problem, e.g., whether or not it is convex. However, the decision making problem in autoscaling tends to be dynamic and uncertain due to the changing QoS and the environment. In contrast, MAs often have broad applicability and less sensitive to the problem's nature. (iii) Additionally, the stochastic nature of MAs allows it to efficiently achieve good coverage in the dynamic and uncertain trade-offs surface of the problem. This improves diversity in the search and hence allows the MAs to efficiently discover better decisions and well-compromised trade-offs. 

Majority of the existing multi-objective optimization work \cite{E3-R-extended}\cite{GA-full-simulation}\cite{2014-eplison-GA-weigh-h-scaling-2014} apply MOGA, most commonly the NSGA-II \cite{nsgaii}, to make decision for autoscaling (we will experimentally compare against them in Section \ref{sec:experiment}). They have ignored QoS interference and thus only considered a limited number of objectives (i.e., up to 4). Unlike their approaches, we have chosen MOACO because (i) it has been shown that MOGA, such as NSGA-II, cannot optimize and make trade-offs for more than 4 objectives \cite{yao}; while our problem needs to handle larger numbers as we consider the trade-offs caused by QoS interference, e.g., we have considered 30 objectives in our experiments. (ii) As discussed in  \cite{yao}, the limitation of MOGA for large number of objectives is due to it needs pareto-dominance to evaluate the overall quality of decisions for all objectives as the algorithm runs; henceforth, causing  the MOGA to obscure and miss important information about the trade-off surface, which restricts its optimality and diversity when the number of objectives increases.  Unlike MOGA, the nature of MOACO allows us to design it in a way that decisions are evaluated against each objective for many single objective optimizations in one run, and thus avoiding the use of pareto-dominance in the optimization. This is achieved by using aggregative heuristics and different pheromone structures for the objectives. Hence, we only need to evaluate the overall quality of decisions for all objectives (i.e., the compromises) after the optimization has been competed. By doing so, the optimization can optimize  and make trade-offs for larger number of objectives while ensuring good diversity. (iii) The sequential pareto-dominance sorting of MOGA can incur large overhead; in contrast, MOACO can gain benefits from parallel programming as each ant works in isolation. (iv) In other problem domains of cloud, e.g., \cite{ant-cloud-vm}, it has been shown that MOACO tends to outperform MOGA.

\textbf{\textit{Thirdly}}, by separating MOACO and the evaluation of decisions' overall quality for all objectives, the MOACO is encouraged to explore more information about the trade-offs surface while saving computational efforts. This design, as shown in \cite{mul-ant-colony}, tends to produce better optimized and diversified trade-off decisions. Hence, we intend to search for well-compromised trade-offs from a set of optimized decisions that exhibit high diversity. Instead of using pure pareto-dominance \cite{Goldberg} to evaluate the overall quality of decisions for all objectives during optimization, we propose a new mechanism, namely \emph{compromise-dominance}, to search well-compromised trade-offs based on the final result of MOACO. Here, we use pareto-dominance \cite{Goldberg} to measure superiority, and a combination of nash-dominance \cite{nash} and the distance of decision to measure fairness. In this way, we aim to achieve a well-balanced improvements for the objectives without explicitly weighting them.

\textbf{\textit{Fourthly}}, we experimentally evaluate our approach using RUBiS \cite{rubis} benchmark and FIFA 98 workload trend \cite{fifa98}. We compare our approach to four widely used approaches for autoscaling: rule-based, heuristic based, randomized and MOGA based.  We have considered four commonly used QoS attributes, these are: Response Time, Throughput, Reliability and Availability. The results suggest that with a large number of objective (i.e., up to 30), our approach produces better trade-offs quality in terms of the numbers of favorable objectives and the extents to which they are optimized; it also produces smaller violation for requirements. Moreover, our approach results in acceptable overheads and has balanced elasticity in terms of over-/under-provision. 

The paper is structured as the following: Section \ref{sec:problem} present a motivating example, assumptions and the formulations of autoscaling problem in the cloud. Section \ref{sec:overview} presents an overview of the autoscaling system. Section \ref{sec:moaco} illustrates the multi-objective ant colony optimization solution. Section \ref{sec:cd} describes the  \emph{compromise-dominance} mechanism for finding well-compromised trade-offs. Section \ref{sec:experiment} presents experiments and evaluations. Section \ref{sec:relatedwork} and \ref{sec:conclusion} present related work and conclusion respectively.


%I wish you the best of success.
%\hfill September 17, 2014

\section{Formal Definition of the Problem Model}
\label{sec:problem}
\subsection{Background and Motivating Example}
Consider, for example, a company called \emph{Rbay} that deploy their services on a public PaaS and IaaS cloud provider, who is currently serving many other cloud consumers. We assume that \emph{Rbay}'s services and numerous other cloud-based services might run in a shared VM or PM, and that these services have different, possibly conflicted QoS and cost objectives, together with time-varying environment changes (e.g., changes in workload). Often, cloud-based services can have multiple replicas for various purposes, e.g., service differentiation and load balancing etc. Therefore we assume that different concrete services {$S_1$, $S_2$, ... $S_i$} might have multiple replicas deployed on different VMs, or even PMs.  In this work, we refer to the replicas of concrete services as \textbf{\emph{service-instances}}: the  \emph{jth} service-instance of the  \emph{ith} concrete service is denoted by $S_{ij}$. Multiple service-instances are  deployed  on  a  cloud  software   stack   running  on  VM, which can be setup using various control knobs. These control knobs are either shared amongst the service-instances (e.g., CPU of the VM) or specific to one service-instance (e.g., threads of a service-instance). Suppose that, from time to time, the QoS or cost requirements of  \emph{Rbay}'s service-instances are violated, autoscaling system needs to continually find the best autoscaling decision\textemdash w.r.t. the QoS and cost objectives\textemdash as to the amount of scaling that should be applied to the necessary control knobs. Such decision making needs to consider the dynamic, uncertainty and trade-offs related to the conflicted objectives, e.g., throughput and cost.

Now, suppose that the autoscaling system has decided to improve the throughput of \emph{Rbay}'s service-instance $S_{ij}$ by provisioning more memory to the underlying VM. Such a decision might not be an issue when the contention is light.  However, as the provision increases, eventually it will result in throughput degradation to the other service-instances on the co-hosted VMs, leading to dynamic QoS interference \cite{2013-JRAO-most-closest-work-2013}\cite{software-RP-two-loops}\cite{qcloud}. The same issue applies when we increase the number of service threads for a service-instance, where the co-located service-instances on the same VM might be interfered \cite{software-RP-two-loops}\cite{2014-decision-tree-software-CP-2014}. These phenomenons imply that there are trade-offs between the throughput of $S_{ij}$ and those of the other service-instances, which might be owned by different cloud consumers. It becomes  more complex when we need to consider trade-offs between conflicted objectives, e.g., the throughput and cost of $S_{ij}$. All these facts can lead to a large number of dependent objectives in a decision making process (i.e., more than 4). Since it is often too expensive to completely eliminate QoS interference \cite{qcloud}, we aim to optimize the services' objectives till the point where interference becomes significant, and then mitigate the effects of QoS interference by making well-compromised trade-offs.

However, well-compromised trade-offs cannot be guaranteed by purely existing pareto-dominance based approaches \cite{Goldberg} (we present the definition of pareto-dominance in Section \ref{sec:cd}). Given the large number of dependent objectives caused by QoS interference, quantifying compromise in the trade-offs purely based on pareto-dominance can lead to a large number of trade-off decisions, which also contain the ones that have imbalanced improvements. Suppose that the autoscaling system has reached two decision \emph{A} and \emph{B} where \emph{A} leads to 9 significantly better objectives than \emph{B}; while \emph{B} can only lead to one slightly better objective than \emph{A}. These two decisions are regard as indifferentiable in the sense of pareto-dominance. Assuming that both decisions satisfy all the requirements constraints, it is generally the case that \emph{A} is more preferable than \emph{B}. However, since they are equivalent in pareto-dominance, \emph{B} can be selected instead of \emph{A}, which results in badly compromised trade-offs.

\subsection{Cloud Primitives}
In this work, we term both control knobs and environment conditions in the cloud as \textbf{\textit{cloud primitives}}, which serve as the fundamental inputs of a QoS model. Without the lose of generality, we further decompose the notion of cloud primitives into two major domains: these are \textbf{\textit{Control Primitive (CP)}} and \textbf{\textit{Environmental Primitive (EP)}}. Control Primitives are the internal control knobs  and  can  be  either  software  or hardware, which can be tuned to support QoS. At the PaaS layer, software control primitives are the key software configurations in cloud; such as the number of threads in thread pool of service/application, the buffer size and load balancing policies etc. The hardware control primitives are computational resources, such as CPU and memory at the IaaS layer. In particular, it is non-trivial to consider software control primitives when autoscaling in the cloud as they have been shown to be important features for QoS  \cite{2013-JRAO-most-closest-work-2013}\cite{software-RP-two-loops}\cite{2014-decision-tree-software-CP-2014}. On the other hand, Environmental Primitives refer to the external stimulus that cause dynamics and uncertainties in the cloud. These, for examples, can be the workload and unpredictable incoming data etc.

\subsection{Objective Model}

The basic notations are explained in Table \ref{table:notations}. Formally, the  QoS objective model at the \emph{tth} sampling interval is:
\begin{equation}
\label{eq:qos}
\mathit{QoS}_k^{\mathit{ij}}(t)=f_k^{\mathit{ij}}(\mathit{SP}_k^{\mathit{ij}}(t),\delta )
\end{equation}
where the selected primitives input matrix $SP_k^{ij}(t)$ is:
\begin{equation}
\label{eq:sp}
\mathit{SP}_k^{\mathit{ij}}(t) =
 \begin{pmatrix}
  \mathit{CP}_a^{\mathit{xy}}(t) & \cdots & \mathit{EP}_b^{\mathit{mn}}(t-1)& \cdots \\
  \vdots  & \ddots  & \vdots & \ddots \\
  \mathit{CP}_a^{\mathit{xy}}(t-q+1) & \cdots & \mathit{EP}_b^{\mathit{mn}}(t-q) & \cdots \ 
 \end{pmatrix}
\end{equation}
whereby \emph{q} is the number of order. In this work, we dynamically create and update $SP_k^{ij}(t)$   and the function $f_k^{ij}$  using the online QoS modeling approach in our prior work \cite{Chen:2013}\cite{Chen:2014:ucc}. Here, we update $SP_k^{ij}(t)$  to contain only the most significant cloud primitives that can influence the QoS, including those that cause considerably high level of QoS interference. This is achieved by using \textit{Symmetric Uncertainty} \cite{Witten:2005}: a metric that quantifies the relevance between two time-series data, which in our case are QoS and cloud primitives. The primitives, which provide the most relevant information to a QoS attribute while causing minimal redundancy to other already selected primitives, are the ones that we are seeking. On the other hand, $f_k^{ij}$ and the required historical data points are determined by a set of machine learning algorithms \cite{Chen:2013}\cite{Chen:2014:ucc}. 

The total cost model for  $S_{ij}$  can be represented as: 
\begin{equation}
\label{eq:cost}
\mathit{Cost}^{\mathit{ij}}=\sum _{a=1}^n\mathit{CP}_a^{\mathit{ij}}(t) \times \mathit{P_a}
\end{equation}
where \emph{n} is the total number of control primitive type that used by service-instance  $S_{ij}$ to supports its QoS attributes.
\begin{table}[t!]
\caption{The Basic Notations of Autoscaling Decision Making Problem.}
\label{table:notations}
\begin{tabularx}{\columnwidth}{|P{1.1cm}|X|}
\hline \hline

$QoS_k^{ij}(t)$ & The \emph{kth} QoS attribute of $S_{ij}$, and its value (e.g., mean response time) at interval \emph{t}.
\\ \hline
$f_k^{ij}$  &  The QoS function for the \emph{kth} QoS attribute of $S_{ij}$.
\\ \hline
$SP_k^{ij}(t)$   &  The selected primitives matrix of $S_{ij}$ at \emph{t}, its column contains the most relevant and significant inputs for the QoS, including the primitives that tend to directly influence the QoS (e.g., the threads of the corresponding service-instance); and the primitives that belong to the co-located service-instances and the co-hosted VMs. The row indicates the number of order, which represents how many historical data points need to be used as inputs for improving model accuracy.
\\ \hline
$\delta$  &  Any other inputs, e.g., historical time-series QoS points and tuning variables etc., that improve model accuracy.
\\ \hline
$CP_a^{ij}(t)$  &  The value of the \emph{ath} control primitive for $S_{ij}$ at interval  \emph{t}, e.g., CPU, memory and thread etc.
\\ \hline
 $\mathit{\emph{\scalebox{1}{EP}}_b^{mn}\emph{\scalebox{1}{(t-1)}}}$  &  The value of the \emph{bth} environmental primitive for $S_{mn}$ at interval \emph{t-1}, e.g., workload etc.
\\ \hline
$P_a$  &  The price per unit of the \emph{ath} control primitive for a service-instance.
\\ \hline
\emph{d}  &  An autoscaling decision consists of newly configured values at \emph{t}, i.e., $d=\langle CP_1^{11}(t), CP_2^{11}(t), ... CP_a^{ij}(t)\rangle$.
\\ \hline
$O_o(t)$   & The \emph{oth} dependent objective in a region. It can be either QoS (\ref{eq:qos}) or cost (\ref{eq:cost}) for the same or different services.
\\ \hline \hline
\end{tabularx}
\end{table}

To reach a trade-offs decision for autoscaling, particularly in the presence of the objective-dependency at both intra- and inter-services level, we dynamically cluster the objectives into different regions in terms of whether they have the common inputs that are parts of the decision. If these common inputs exist, it means that the objectives are dependent (i.e., harmonic or conflicted) and can be affected differently by the same decision; hence, they need to be considered in conjunction with each other in the decisions making process. On the other hand, the objectives, which are independent, are omitted from the same decision making process as they can benefit nothing but generate overhead. Please refer to our prior  work  \cite{Chen:2014}  for  detailed  specification  of  the  region clustering approach. As a result, autoscaling in the cloud needs to optimize multiple independent regions, each of which contains a different set of objectives. 

For each region, our ultimate goal is to produce an autoscaling decision \emph{d} that uses the minimal costs to achieve the best possible QoSs, shown as the following:
\begin{equation}
\label{eq:objective}
\mathit{Maximize} \ or \ \mathit{Minimize} \  \big \langle O_1(t),O_2(t)...O_o(t) \big \rangle
\end{equation}
subject to
\begin{equation}
\label{eq:condition1}
 \forall QoS_k^{ij}(t) \succeq SLA_k^{ij}
 \end{equation}
 \begin{equation}
 \label{eq:condition2}
 \forall Cost^{ij}(t) \leq Budget^{ij}
 \end{equation}
 \begin{equation}
 \label{eq:condition3}
 min_a^{ij} \leq  \forall CP_a^{ij}(t)  \leq  max_a^{ij}
\end{equation}
whereby (\ref{eq:condition1}) states that any QoS attribute should meet its Service Level Agreement (SLA). (\ref{eq:condition2}) denotes that the cost of each service-instance should not exceed its budget requirement on a VM. Finally, (\ref{eq:condition3}) represents that the possible configured values of control primitives must be selected from a given range of the underlying hardware or software. $min_a^{ij}$ and   $max_a^{ij}$ are the thresholds to control the range of possible configured values, and they are dynamically updated online, as we will see in Section \ref{sec:overview}. 

It is obvious that, by omitting any predefined weights of objectives in (\ref{eq:objective}), we render the problem as a discrete multi-objective optimization problem, which involves multiple trade-offs and is usually NP-hard \cite{Antonescu13dynamicsla}\cite{E3-R-extended}\cite{GA-full-simulation}.

\section{Overview of the Autoscaling System}
\label{sec:overview}
\begin{figure}[t!]
\centering
\includegraphics[width=9cm]{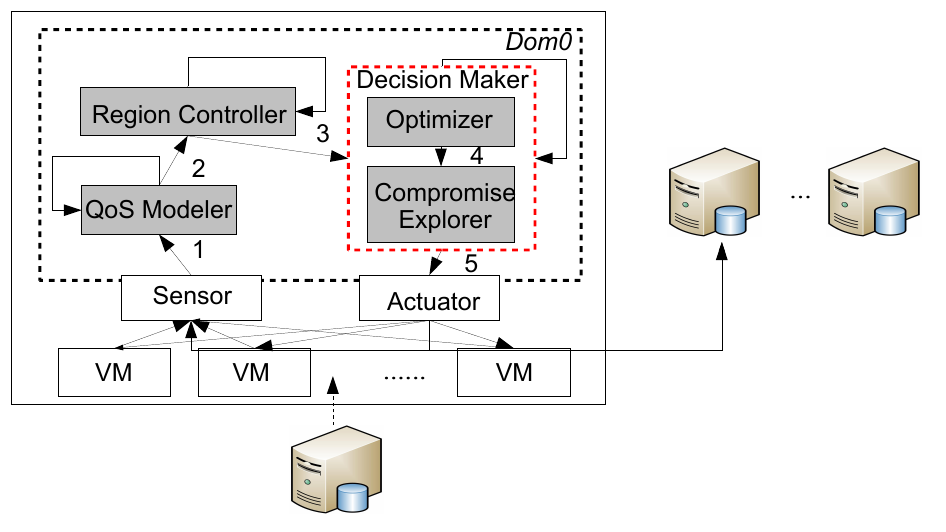}
\caption{The architecture of autoscaling system.}
\label{fig:arch}
\end{figure}

To enable self-adaptive decision making for autoscaling in the  cloud,  we   have   designed   and   implemented   an autoscaling system using decentralized, multiple feedback loops, which run on the root domain of each PM, as shown in Fig. \ref{fig:arch}. The components in our system, except \textit{QoS Modeler}, are triggered when it detects violations of the requirements, i.e., violations of SLA, and utilization constraints in case of over-provision. The sensors on a PM does not only sense data, but also the QoS models from other PMs. This is because in some cases, a cloud-based service can be functionally dependent on services running on the other PMs, thus creating the chances for objective-dependency.

To build the QoS model in (\ref{eq:qos}), we have realized self-adaptive and online QoS modeling in the \textit{QoS Modeler}, as described in our prior work \cite{Chen:2013}\cite{Chen:2014:ucc}. Subsequently, in the self-adaptive \textit{Region Controller}, the most up-to-date QoS models and the cost models are dynamically clustered into regions; more details can be found in our prior work \cite{Chen:2014}.  
 
Once the regions of objectives are identified, the focus of this work is to adaptively produce decision that achieves well-compromised trade-offs with respect to the  objectives in each region. To this end, we design the self-adaptive \textit{Decision Maker} component. Specifically, one of the subcomponents of \textit{Decision Maker} is the \textit{Optimizer} component, which leverages on MOACO to search and optimize for the possible set of trade-offs decisions. Next, another subcomponent, namely \textit{Compromise Explorer},  extracts the decisions that achieve well-compromised trade-offs from the result of Optimizer for autoscaling in the cloud. Theses two subcomponents are specified in Section \ref{sec:moaco} and \ref{sec:cd} respectively. To control the diversity of possible decisions, the possible configured values for each control primitive of a service-instance are bounded within a range, as mentioned in (\ref{eq:condition1}). The lower bound is set to the maximum of a predefined value and the latest observed one. On the other hand, after the upper bound is set as an initial value, it is then dynamically adjusted based on the newly decided value and the latest observed one, i.e., it is increased by \emph{k}\% if both values converge to the upper bound; likewise, it is decreased by \emph{k}\% if both values diverge from the upper bound.

We consider both vertical scaling and horizontal scaling in the actuators. The former refers to change the configurations and provision of control primitives within a PM; the later refers to boots up/shutdown VMs on the other PMs via migration or replication. In our system, vertical scaling always takes higher priority, providing that modern hypervisors (e.g., Xen \cite{xen}) achieve dynamic vertical scaling with negligible overheads. The resources on a PM are provisioned to the VMs in a first-come-first-serve basis. The horizontal scaling, on the other hand, is only triggered when the resources of the PM tends to be exhausted, i.e., when the total upper bounds of all co-hosted VMs for a resource type exceeds the PM's capacity. Likewise, a VM is removed when its provisions and utilizations  for all resource types are below thresholds.

\section{Searching and Optimizing Trade-offs Decisions For Autoscaling in the Cloud}
\label{sec:moaco}
Recall that for each region, our aim is to optimize (\ref{eq:objective}) subject to the requirements and constraints in (\ref{eq:condition1})-(\ref{eq:condition3}). To this end, we follow the multi-objective ant system described by \cite{mul-ant-colony}, in which there are one colony and \emph{m} pheromone structures where \emph{m} is the number of objectives being optimized. The MOACO relies on probabilistic search-based optimization that assumes a fixed number of iterations. In each iteration, the ants select different QoS or cost objectives to optimize for. By the end of an iteration, each ant produces an autoscaling decision containing the selected configured values for those cloud control primitives that are inputs of the objectives in (\ref{eq:objective}). This is achieved by the use of a probabilistic rule, which expresses the desirability for an ant to choose a particular value for a control primitive. This rule is based on the information about the current pheromone trail, which drives the ants to search better decisions for a particular objective that is selected; and an aggregative heuristic that guides the ants toward choosing better overall decisions with respect to all objectives. Hence, the higher the amount of pheromone and heuristic information is associated with a particular value of a control primitive, the higher the probability is that an ant will choose it.  This stochastic nature of the algorithm allows the ants to explore largely diversified decisions as the search proceeds. 

%Henceforth, it provides more alternatives from the trade-off surface than approaches that based on weighted-sum of objectives. 

\subsection{Probabilistic Rule}

Suppose that an ant selects the \emph{oth} QoS or cost objective to optimize; the probability for selecting the \emph{xth} configured value of the \emph{ath} control primitive is defined by:
\begin{equation}
\label{eq:probability}
p_{x,a,o}=\frac{(\tau _{x,a,o})^{\alpha }\times (\eta _{x,a})^{\beta }}{\sum _{y\in S}^{}(\tau _{y,a,o})^{\alpha }\times
(\eta _{y,a})^{\beta }}
\end{equation}
\noindent whereby \emph{S} denotes the set of possible configured values for the \emph{ath} control primitives; $\tau _{x,a,o}$ is the pheromone for the \emph{xth} configured value of the \emph{ath} control primitives when optimizing the \emph{oth} objective. $\eta _{x,a}$ is the heuristic factor for the \emph{xth} configured value of the \emph{ath} control primitive. $\alpha$ and $\beta$ are two parameters that determine their relative importance. It is worth noting that for each ant, the control primitives, which we need to find configured values for, are not restricted to the inputs of the \emph{oth} objective function, but also include those of the other QoS or cost objectives in (\ref{eq:objective}).

\subsection{Heuristic Factor}

Instead of aggregating the objectives to be optimized, we aggregate the heuristic information to favor the decisions that tend to improve the overall quality of all objectives. In this way, we aim to handle large number of objective while do not require to specify weights on objectives. To this end, we leverage on the normalized, scalar-valued difference between the total improvement and the total degradation for all objectives. This is achieved by comparing the outputs when using a newly configured value to that of the original value, formally expressed as:
\begin{equation}
\label{eq:heuristic}
\eta _{x,a}=\begin{cases}\frac{\sum _{o=1}^mI_{x,a,o}}{1+\sum _{o=1}^mD_{x,a,o}}
& \mathit{if}\sum_{}^{}I_{x,a,o}\neq 0 \\ 
\frac{\eta _{x,a}^{\mathit{min}}}{1+\sum_{o=1}^mD_{x,a,o}}
& \mathit{otherwise}\end{cases}
\end{equation}
\noindent whereby, for all \emph{m} objectives that need to be optimized, $\sum _{o=1}^mI_{x,a,o}$ is the total improvement over the current setup when using the \emph{xth} configured value of the \emph{ath} control primitive in the decision. Likewise, the total degradation in contrast to the current setup is denoted by $\sum _{o=1}^mD_{x,a,o}$. To prevent zero heuristic factor in case where the configured values cannot improve any objectives, we use the minimum non-zero heuristic over all possible configured values, denoted by $\eta _{x,a}^{min}$ , as the initial value. In this way, even though the ants select different single objectives to optimize for, the heuristic still ensure that the configured values, which lead to overall better decisions, are relatively more attractive. 

%It has been shown that these operators promotes better coverage for searching possible trade-offs decisions \cite{4410319}. 

%To improve numeric stability, we normalize the improvement and degradation of an objective by dividing them to a reference point, which is the output of the said objective when using the original configured values.

\subsection{Pheromone Update}

After all ants complete the search in an iteration, the pheromone trails need to be updated in order to help guiding the search towards better decisions. Unlike the heuristic information, the pheromone is designed to favor the decisions that improve each objective individually. As a result, the pheromone for a particular configured value of a control primitive is specific to an objective. Each pheromone trail is updated by using the rule below:
\begin{equation}
\label{eq:pheromone}
\tau _{a,x,o}=(1-\rho )\times \tau _{a,x,o}+\Delta \tau _o^{\mathit{best}}
\end{equation}
\noindent where $\rho$ ($0< \rho <1$) is a constant that simulates the evaporation of pheromone trails, it determines the speed of evaporation\textemdash a larger value implies faster evaporation. Thus, the corresponding configured value becomes unattractive quicker. $\Delta \tau_o^{\mathit{best}}$ is a factor that deposits the pheromone for some favorable decisions. In this work, we follow the MAX-MIN Ant System \cite{mul-ant-colony} in which only the configured values that belongs to the iteration's best decision can deposit the pheromone, as defend by:
\begin{equation}
\label{eq:delta-pheromone}
\Delta \tau_o^{\mathit{best}}=\begin{cases}\frac{1}{1 + h(d_o^{\mathit{best}})^{-1} - h(d_o^{global-best})^{-1}}
& \mathit{if} \ x\in d_o^{\mathit{best}},\mathit{max} \ h\hfill\null
\\ 
\frac{1}{1 + h(d_o^{\mathit{best}}) - h(d_o^{global-best}) }
& \mathit{if} \ x\in d_o^{\mathit{best}},\mathit{min} \ h\hfill\null
\\  
0 
& \mathit{otherwise}\end{cases}
\end{equation}
\noindent where $d_o^{best}$ is the best decision for the \emph{oth} objective at the current iteration, and $d_o^{global-best}$ˆ' denotes the best ever decision for the \emph{same} objective; \emph{h} is the corresponding objective function. As such, the configured values in the best decisions would become more attractive; whereas the others, which are not part of the best decisions, will lose pheromone based on the speed of evaporation. In addition, by introducing the best ever decision in the update, we force the search towards the optimal decision for an objective. It is easy to see that, by incorporating the heuristic information and pheromones, the MOACO favors the decisions that do not only benefit all objectives, but also tend to improve each individual objective as much as possible. In this way, the harmonic objectives can be continually optimized in parallel till they reach the point where trade-off needs to be made; while the conflicted objectives would be forced to make trade-offs from the beginning. Consequently, the MOACO is able to produce higher diversity in the trade-off decisions.

Finally, given the fact that only the iterations' best decisions are allowed to deposit the pheromone, the ants may always conclude in the same or similar decisions, which causes the search to be tracked in local spaces. To resolve this issue, we leverage on the solution as used by the MAX-MIN Ant System \cite{mul-ant-colony}, where the pheromone for the  \emph{oth} objective are bounded within a given range,  denoted as $\tau_o^{max}$ and $\tau_o^{min}$ . By the end of an iteration, the bounds are updated using the iteration's best decision:
\begin{equation}
\label{eq:pheromone-update1}
 \tau
_o^{\mathit{max}}=\begin{cases}\frac{1}{h(d_o^{\mathit{best}})^{-1} \times (1-\rho)}
&  if \ \mathit{max} \ h\hfill\null
\\ 
\frac{1}{h(d_o^{\mathit{best}})  \times (1-\rho)}
&  if \ \mathit{min} \ h\hfill\null
\end{cases}
\end{equation}
\begin{equation}
\label{eq:pheromone-update2}
\tau_o^{min} = v \times \tau_o^{max}
\end{equation}
\noindent where  \emph{h} is the corresponding objective function.  \emph{v} is a factor that controls the length of the range.

\subsection{Workflow of MOACO}

The workflow of MOACO can be described as the following:

\begin{itemize}[leftmargin=.5cm]

\item \textbf{\textit{Step 1:}} setup the configurations, e.g., number of iterations (\emph{maxIteration}), number of ants (\emph{maxAnt}) and the number of runs for an ant to find satisfactory decision (\emph{maxRun}).
\item \textbf{\textit{Step 2:}} compute the heuristic information (\ref{eq:heuristic}) and initialize the pheromone trails with an identical value.
\item \textbf{\textit{Step 3:}} each ant simultaneously selects an objective to optimize. To form a decision, an ant chooses the configured value of each related control primitive via probabilistic rule (\ref{eq:probability}). If no satisfactory decisions found (i.e.,  those that have no violations of requirements for all objectives), the ant repeats till it reaches \emph{maxRun} and returns the best decision for the selected objective.
\item \textbf{\textit{Step 4:}} upon completion of an ant, it stores the identified decision. If such decision is the best for the selected objective in the current iteration, the ant updates the objective's local best decision.
\item \textbf{\textit{Step 5:}}  after all ants produce decisions for all the objectives, the best global decision for an objective, which results in the best value so far, is updated if a better decision found.
\item \textbf{\textit{Step 6:}} update pheromone bounds using (\ref{eq:pheromone-update1}) and (\ref{eq:pheromone-update2}), and the pheromone trail of each possible configured value for an objective using (\ref{eq:pheromone}) and (\ref{eq:delta-pheromone}).
\item \textbf{\textit{Step 7:}} the search terminates when it reaches its maximum iterations, and returns all the decisions identified. Otherwise it repeats from Step 3.

\end{itemize}

\section{Identifying Well-Compromised Trade-offs for Autoscaling}
\label{sec:cd}
It is clear that MOACO is able to search the possible trade-off decisions for autoscaling in the cloud; however,  it does not cater for the dynamic and uncertainty of the good compromises in the trade-offs. In this section, we present a simple but efficient mechanism, namely \emph{compromise-dominance}, to adaptively find the decisions that achieve well-compromised trade-offs from the result of MOACO. Specifically, our \emph{compromise-dominance} consists of two phases: superiority phase and fairness phase.
%given the large amount of decisions produced,

\subsection{Superiority Phase}

The first phase in our \emph{compromise-dominance} mechanism is to ensure the superior decisions, which are clearly more favorable than the others. To achieve this, we use the well-known principle of pareto-dominance \cite{Goldberg}:
\begin{quote}
\textbf{Pareto-Dominance:} \textit{A decision $d_1$ pareto-dominates another $d_2$, if and only if, (i) all the objective results achieved by $d_1$ are better than or equivalent to those achieved by $d_2$; and (ii) the result of at least one objective achieved by $d_1$ is better than the result of the same objective achieved by $d_2$.}
\end{quote}
It is easy too see that if a decision pareto-dominates another, then it is better than another in terms of the quality of every individual objective and the overall quality for all objectives. In such context, the decisions, which are not pareto-dominated by any others, are called non-pareto-dominated decisions. These decisions are pareto optimal in case no objective can be further improved without making the other objectives worse off. Our aim in this phase is to identify the non-pareto- dominated objectives. If they do not exist, we use the decisions that being pareto-dominated the least.

\subsection{Fairness Phase}

After the superior decisions are determined, the second phase aims to ensure the fairness in a decision. That is to say, we are interested in making the trade-offs well-balanced with respect to all objectives. To this end, we leverage on nash-dominance \cite{nash}:
\begin{quote}
\textbf{Nash-Dominance:} \textit{A decision $d_1$ nash-dominates another $d_2$ , if and only if, there are less objectives that can improve their results by switching from $d_1$ to $d_2$ than vice-versa.}
\end{quote}
If a decision nash-dominates another, it means that it is more fair with respect to all objectives, and thus more stable. In particular, the decisions, which are not nash- dominated by others, are called non-nash-dominated decisions. As proven in \cite{nash}, a non-nash-dominated decision reaches Nash Equilibrium where no objective can be further improved without changing the results of other objectives. It has been shown that, Nash Equilibrium is the most fair state for all objectives in the sense that it exhibits fair competition, or compromise \cite{nash}. Here, our aim is to identify the non-nash-dominated objectives; or those that being nash-dominated the least if there is no non-nash-dominated objectives.

However, nash-dominance tends to be limited in reducing the number of decisions when the number of dependent objectives is small, e.g., less than 4 objectives.  To this end, we use an additional metric, namely \textit{distance of decision}, to select well-compromised trade-offs under those cases. Concretely, we select the best value of each objective from all the decisions identified; these values form a reference point. We then calculate the normalized \textit{Euclidean Distance} of the result, which is achieved by each decision, to this reference point. The decision(s), which leads to result that has the minimal distance, is the one(s) that we are seeking. 

\subsection{Workflow of Compromise-Dominance}

We now explain the workflow of the compromise-dominance mechanism using the following steps:

\begin{itemize}[leftmargin=.5cm]
\item \textbf{\textit{Step 1:}} find the satisfactory decision from the set of decisions identified by MOACO. If it fails to do so, it selects the decisions that result in the least number of violated requirements.  This is because in some cases, the violations are inevitable outcomes due to, e.g., heavy conflicts amongst the objectives and/or improper settings of the requirements. 
\item \textbf{\textit{Step 2:}} rank each decision in the set based on the number of other decisions that pareto-dominate it. Smaller number represent higher rank of a decision. Thus, we select a subset of decisions that is ranked the highest in terms of pareto-dominance.
\item \textbf{\textit{Step 3:}} rank the reduced set from Step 2 using nash-dominance. The less a decision is nash-dominated, the higher the rank is. Likewise, we select a set of decisions that is  ranked as the highest under nash-dominance.
\item \textbf{\textit{Step 4:}} calculate the reference point. Based on the reduced set from Step 3, search the decisions that leads to result that has the  smallest distance to that point.
\item \textbf{\textit{Step 5:}} randomly select one decision from the final set.

\end{itemize}

\section{Experiments and Evaluations}
\label{sec:experiment}
We integrate our MOACO and the \emph{compromise-dominance} (CD) mechanism, denoted as MOACO-CD. To evaluate the proposed approach, we have conducted various quantitative experiments. The primary goal of these experiments is to validate the effectiveness of our approach against other state-of-the-art autoscaling approaches in cloud, these are:

\begin{itemize}[leftmargin=.5cm]

\item \textbf{\emph{RULE}} - A conventional rule-based autoscaling approach that makes decisions using predefined \emph{if-conditions-then-action} mapping, e.g., \cite{Compsac_2010_I_Brandic}\cite{scale-rule-based}. This approach does not require explicit QoS model as the QoS of a service-instance is assumed to be sensitive to its own control primitives only, e.g., the CPU and thread of the said service-instance. Specifically, violations of QoS would increase all the relevant control primitives to the next higher value; while low utilization would decrease them to the next lower value.

\item \textbf{\emph{HILL}} - A more sophisticated autoscaling approach that relies on our QoS modeling \cite{Chen:2013}\cite{Chen:2014:ucc} and region controlling technique \cite{Chen:2014}, but the decision making process leverages on a weighted-sum formulation of all the dependent objectives, e.g., \cite{2014-profiling-decision-tree-scaling-2014}\cite{smart2004}. Here, the approach leverages on greedy and heuristic based solution: the random-restart hill-climbing algorithm for optimization, in which it starts with an arbitrary decision, then attempts to find a better decision by incrementally and independently changing the values of each control primitives in the models. The algorithm terminates when a maximum iteration has been reached. The best decision, in terms of the weighted-sum formulation, is returned.

\item \textbf{\emph{RANDOM}} - Another autoscaling approach that is similar to HILL, but instead of using hill-climbing, a random optimization algorithm is applied (e.g.,  \cite{Chen:2014}). This algorithm randomly changes the values of each related control primitive, and terminates when it reaches a maximum number of iterations. The best decision is selected as indicated by the weighted-sum formulation.

\item \textbf{\emph{MOGA}} - A  most commonly used multi-objective genetic algorithm derived from NSGA-II, e.g., \cite{E3-R-extended} \cite{GA-full-simulation} \cite{2014-eplison-GA-weigh-h-scaling-2014}. We have also designed MOGA to benefit from our QoS modeling \cite{Chen:2013}\cite{Chen:2014:ucc} and region controlling technique  \cite{Chen:2014}. In addition, we configure the optimal population size and number of iterations through careful profiling on our testbed. 

\end{itemize}

Notably, we have configured the approaches to use the identical number of global iterations for the worst case. However, to prevent them from completing with arbitrary latency, we have set a running time threshold (i.e., 75s), which forces the algorithms to terminate and return the best decision found. For HILL and RANDOM, we normalize each objective's result in the weighted-sum of objectives and set all the weights to 1. We use the following 5 criteria to quantify the comparisons:

 \textbf{\emph{Coverage of two approaches (C-metric)}} \cite{797969} - this metric performs pairwise comparison to measure the comparative quality of trade-offs achieved by two approaches. It is calculated using the number of (relatively) better objectives achieved by one approach, divided by the total number of considered objectives. Formally, the C-metric is defined as:
\begin{equation}
\label{eq:cmetric}
C(A,B)=\frac{\left|r_{o,a}\in A:r_{o,b}\in
B,r_{o,a} \succeq r_{o,b}\right|}{m}
\end{equation}
whereby \emph{A} and \emph{B} represent two approaches and their corresponding sets of average objective results for all intervals that are being considered. $r_{o,a}$ and $r_{o,b}$ are the average results of the \emph{oth} objective, as achieved by the two approaches; these average results are calculated by averaging the objective values for \emph{n} intervals, as denoted by $ r_{o,a}=\frac 1 n\times \sum _{i=1}^nr_{i,o,a}$. \emph{m} is the total number of objectives that we consider. $\left|r_{o,a}\in A:r_{o,b} \in B,r_{o,a} \succeq r_{o,b}\right|$ counts how many objective results achieved by \emph{A} are better than those achieved by \emph{B}. Intuitively, the C-metric is an effective method to quantify the quality of trade-offs with respect to the number of the favorable objectives. The greater the value is, the better the approach is. $C(A,B)=1$ means that the results of all objectives achieved by \emph{A} are better than those achieved by \emph{B}.
%\;\ \mathit{s.t.}, \ r_{o,a}=\frac 1 n\times \sum _{i=1}^nr_{i,o,a}
 
\textbf{\emph{Generational Distance (G-Distance)}} \cite{gdistance} - this is another intuitive metric that measures the quality of trade-offs. Unlike the C-metric, G-Distance focuses on the generational extents to which the objectives are optimized as achieved by an approach. Formally, it is calculated by:
\begin{equation}
\label{eq:gdistance}
G-Distance=\sqrt{\sum _{o=1}^{m}(\frac 1{r_o^{\mathit{max}}}\times ((\frac 1 n\times
\sum _{i=1}^nr_{i,o,a})-r_o^{\mathit{best}}))^2}
\end{equation}
where $r_{i,o,a}$ is the result of the \emph{oth} objective at the \emph{ith} interval, as achieved by the \emph{ath} approach.  $r_{o}^{best}$ and $r_{o}^{max}$ are the best and the max average result (over all approaches) for the \emph{oth} objective respectively. Smaller value of G-Distance means better result. The remaining notations are the same as (\ref{eq:cmetric}).

 \textbf{\emph{Violations of Requirements}} - for each approach, we
measure the extent to which the requirements (i.e., SLA or budget) of an objective are violated, as defined in:
\begin{equation}
\label{eq:sla}
\frac{100} n\times \sum
_{i=1}^nv_i\ \ \mathit{s.t.},\ v_i=\begin{cases}\frac{\left|r_{i,o,a}-t_o\right|}{t_o}
& \mathit{if} \ t_o \succeq r_{i,o,a}
\\ 
0 & \mathit{otherwise}\hfill\null
\end{cases}
\end{equation}
whereby $v_i$ is the extent of violation at the \emph{ith} interval; $t_o$ is the requirement threshold for the \emph{oth }objective, i.e.,  SLA or budget; and  \emph{n} is the total number of intervals.

 \textbf{\emph{Over- and Under-Provisioning}} - for each approach, we quantify over-/under-provision by means of the average difference between the provision and demand for each control primitive type. The over-provision is calculated as:
\begin{equation}
\label{eq:provision}
\frac{100}{m\times n}\times \sum _{j=1}^m\sum
_{i=1}^nU_{i,j}\ \mathit{s.t.},\ U_{i,j}=\ \left\{\begin{matrix}\frac{\left|u_{i,j}-u'_{i,j}\right|}{u'_{i,j}}\ \ \ \mathit{if} \ u_{i,j}>u'_{i,j}\\\  0\ \ \ \ \ \ \ \  \  \ \ \ \ \ \ \mathit{otherwise}\hfill\null
\end{matrix}\right.
\end{equation}
where $u_{i,j}$ is the provision of the \emph{ith} control primitive for the \emph{jth} VM (for hardware control primitives) or  the \emph{jth} service-instance (for software control primitives) on a PM. \emph{n} and \emph{m} are respectively the number of intervals and VMs/service-instances.  $u'_{i,j}$ is the corresponding demand using the highest possible value that we have observed. The calculation of under-provision can be similarly applied.

 \textbf{\emph{Overhead}} - We measure the overhead of each approach in terms of the latency in making decisions. Particularly, we report on the results for both the best and worst cases.

\subsection{Experiments Setup}

We conducted experiments on private cloud using a cluster of PMs, each of which has Intel i7 2.8GHz Quad Cores and 4GB RAM. The PMs use Xen v3.0.3 \cite{xen} as the hypervisor and the autoscaling process is running on Dom0. To eliminate the interference caused by Dom0, we allocated one CPU core and 600 MB RAM to it, which tends to be sufficient. Our approach and the other competitors are implemented using Java JDK 1.6. To simulate QoS interference caused by the VMs while not exhausting resources, we run three co-hosted VMs on each PM. Initially, we allocate the same amounts of hardware resources for each of the co-hosted VMs, these are 30\% cap of a dedicated CPU core and 250 MB RAM. All VMs run linux kernel v2.6.16.29.

Our experiments leverage on RUBiS \cite{rubis}, which is a cloud-based application consists of 26 co-located services using the eBay.com model. For simplicity, we have used three RUBiS snapshots, each of which consists of a 2-tiers (i.e., application and database tiers) based RUBiS application. A RUBiS snapshot is deployed with a software stack including linux kernel v2.6.16.29, Tomcat v6.0.28 and MySQL v3.23.58 on each co-hosted VM of the master PM. The snapshots use heterogeneous database volume size ranging from 1GB to 5GB data. We have implemented sensors and actuators on each service-instance/VM for collecting the online data and scaling the control primitives respectively. 

In this work, we have realized vertical scaling actions (a.k.a. scale-up/-down) by using a customized listener on Tomcat and the management module of Xen. As for horizontal scaling actions (a.k.a. scale-in/-out), we leverage on master-salves based replication. Each of the three RUBiS snapshots and its replicas are linked to a dedicated load balancer. Three client emulators are used and they apply read/write pattern to generate requests for each load balancer. To simulate a realistic workload within the capacity of our testbed, we vary the number of clients according to the compressed FIFA98 workload \cite{fifa98}. This setup can generate up to 400 parallel requests, which is large enough to simulate QoS interference.

\subsection{QoS Attributes, Primitives and Configurations}

For the simplicity of exposition, we have selected commonly used QoS attributes and cloud primitives  in the evaluation. In our experiments, we have used identical setups for all approaches. As listed in Table \ref{table:config}, these QoS attributes and primitives are  per-service  except  for  CPU and memory as they are shared on a VM. Table \ref{table:cp-config} shows the configurations for each control primitive type. Scale-out occurs if the summed \emph{max} of CPU or memory for all the co-hosted VM exceeds the PM's capacity. The hardware and software control primitives of a new replica VM and service are set as the initial value \emph{i}. Likewise, scale-in occurs if CPU and memory of a VM are provisioned as \emph{min}, and their utilizations are below \emph{u}. Table \ref{table:sla-budget} illustrates the SLA and budget (per interval on a VM) for each managed service-instance and Table \ref{table:moaco-config} is the configurations for MOACO. By carefully examining the objective-dependency of services based on our prior work \cite{Chen:2013}\cite{Chen:2014:ucc}\cite{Chen:2014}, we intend to manage and autoscale the services that exhibit the most fluctuated performance, and those that are the most likely to lead to the largest number of dependent objectives in a decision process. We have identified two such services on each RUBiS snapshot while leaving the other 24 services as unmanaged, generating interference  only.  All  these  setups  give  us  up  to  30 dependent objectives in one decision making process. 

In each experiment run, the sampling and modeling intervals are both 120s with the total of 70 intervals; and there is one new sample per interval for updating the QoS models. The autoscaling process is triggered when any violations of SLA or low utilization is detected. Given that the QoS modeling requires certain historical data to build the models, we report the achieved QoS and cost of all managed service-instances on the master PM for the rear 50 intervals. Each approach is examined for 10 runs.

\begin{table}[t!]
\centering
  \caption{The Examined QoS Attributes and Primitives.}
\label{table:config} 
\begin{tabularx}{\columnwidth}{|P{0.8cm}P{1.3cm}X|}

\hline \hline
\multicolumn{2}{|c}{\centering QoS and Primitives} &
{\centering Description} \\ 
\hline \hline
\multirow{4}{*}{Output} &
Response Time  (ms) &
The average leaped time between a service-instance receives and replies a request.\\
 &
Throughput  (req/min) &
The average rate of completed requests.\\
 &
 Reliability (\%) &
The percentage of requests that being completed faster than the SLA. (2-4 ms)\\
 &
 Availability (\%) &
The percentage of time that the average response time above a threshold. (4 ms)\\ \hline \hline
 \multirow{3}{*}{CP input} &
 CPU (\%) &
Observed average CPU utilization of a VM.\\
 &
 Memory (MB) &
Observed average Memory utilization of a VM.\\
 &
 Thread (no. of req) &
Observed maximum concurrent threads of a service-instance. (a modified control knob of Tomcat's \textit{maxThread}
property)\\\hline \hline
 EP input &
 Workload (req/min) &
Observed average request rate of a service-instance.\\\hline \hline
\end{tabularx}
\end{table}

\begin{table}[t!]
\centering
  \caption{Configurations for Each Control Primitive Type.}
\label{table:cp-config} 
\begin{tabularx}{\columnwidth}{|P{0.8cm}YYYYYYYP{0.7cm}|}

\hline \hline
 &
 i &
 u &
 step &
 min &
 max &
 t &
 k &
 p \\ 
\hline \hline
CPU &
30\% &
50\% &
1\% &
15\% &
40\% &
70\% &
10\% &
\$0.01 \\ 

Memory &
250MB &
50\% &
5MB &
230MB &
280MB &
70\% &
10\% &
\$0.002 \\ 

Thread &
5 &
50\% &
1 &
4 &
10 &
70\% &
10\% &
\$0.017 \\ 
\hline \hline
\end{tabularx}
\begin{tablenotes}
      \small
      \item  \textit{i = the initial value; u = the lowest possible utilization for triggering autoscaling; step = the margin between two neighbor values; min = the minimum value; max = the maximum value; t = the \% threshold to trigger change of the max value; k = the \% extent to which the max value is changed; p = the price per unit per interval for a service-instance.}
    \end{tablenotes}

\end{table}

\begin{table}[t!]
\centering
  \caption{SLA and Budget for The Managed Service-Instances. }
\label{table:sla-budget} 
\begin{tabularx}{\columnwidth}{|P{0.3cm}P{0.7cm}P{1.3cm}P{1.2cm}P{1cm}P{1.05cm}Y|}

\hline \hline
\multicolumn{2}{|c}{} &
Response Time (ms) &
Throughput (req/min) &
Reliability (\%) &
Availability (\%) &
Cost (\$)   \\ 
\hline \hline
\multirow{2}{*}{VM1} &
Service1 &
2 &
180 &
85 &
90 &
1.2 \\
&
Service2 &
2 &
180 &
85 &
90 &
1.1 \\
\hline
\multirow{2}{*}{VM2} &
Service3 &
3 &
150 &
85 &
90 &
1.17 \\
&
Service4 &
2 &
180 &
85 &
90 &
1.33 \\
\hline
\multirow{2}{*}{VM3} &
Service5 &
4 &
140 &
90 &
85 &
1.02 \\
&
Service6 &
2 &
180 &
90 &
90 &
1.17 \\

\hline \hline
\end{tabularx}
\end{table}

\begin{table}[t!]
\centering
  \caption{Configurations of MOACO. }
\label{table:moaco-config} 
\begin{tabular}{|ccccccc|}

\hline \hline
 $\alpha$  &
  $\beta$ &
  $\rho$ &
  $v$ &
 \emph{maxIteration} &
 \emph{maxAnt}  &
 \emph{maxRun}   \\ 
\hline \hline
4 &
1 &
0.1 &
0.5 &
5 &
150 &
100 \\ 
\hline \hline
\end{tabular}
\end{table}

\subsection{Quality of Trade-offs}

To evaluate the quality of trade-offs achieved by our approach, we leverage on the aforementioned C-metric (\ref{eq:cmetric}) and G-Distance (\ref{eq:gdistance}); the results are plotted in Table \ref{table:quality-trade-offs}. For C-metric, our MOACO-CD is better than MOGA as the later is limited in optimizing and making trade-off for a large number of objectives; it also does not consider well-compromised trade-offs.  MOACO-CD is superior to RULE, which does not allow explicit optimization and trade-off. Finally, MOACO-CD is also better than HILL and RANDOM, because the weighted-sum of objectives in these two has greatly restricted their search into local areas of the search space, henceforth they tend to be limited in improving the diversity of trade-offs decisions. As a result, we can conclude that our MOACO-CD is the best according to C-metric, meaning that it has the best quality of trade-offs in terms of the number of the favorable objectives.

\begin{table}[t!]
\centering
  \caption{Quality of Trade-offs. (The Best is Highlighted in Bold)}
\label{table:quality-trade-offs} 
\begin{tabularx}{\columnwidth}{|ccYP{1.2cm}YY|}
\hline \hline
\multicolumn{6}{|c|}{Pairwise Comparison on C-metric} \\ 
\hline \hline
\multicolumn{6}{|c|}{C(\textbf {MOACO-CD},MOGA):C(MOGA,\textbf {MOACO-CD})=0.8:0.2} \\
\multicolumn{6}{|c|}{C(\textbf {MOACO-CD},RULE):C(RULE ,\textbf {MOACO-CD})=0.73:0.27}\\
\multicolumn{6}{|c|}{C(\textbf {MOACO-CD},HILL):C(HILL,\textbf { MOACO-CD})=0.8:0.2}\\
 \multicolumn{6}{|c|}{C(\textbf {MOACO-CD},RANDOM):C(RANDOM,\textbf {MOACO-CD})=0.73:0.27}\\
\hline \hline
&
MOACO-CD &
MOGA &
RANDOM &
RULE &
HILL \\ \hline \hline
G-Distance &
\bfseries 0.4071 &
1.2707 &
0.9407 &
1.5892 &
1.6958 \\

\hline \hline

\end{tabularx}
\end{table}

\begin{figure}[t!]
\centering
\includegraphics[width=\linewidth]{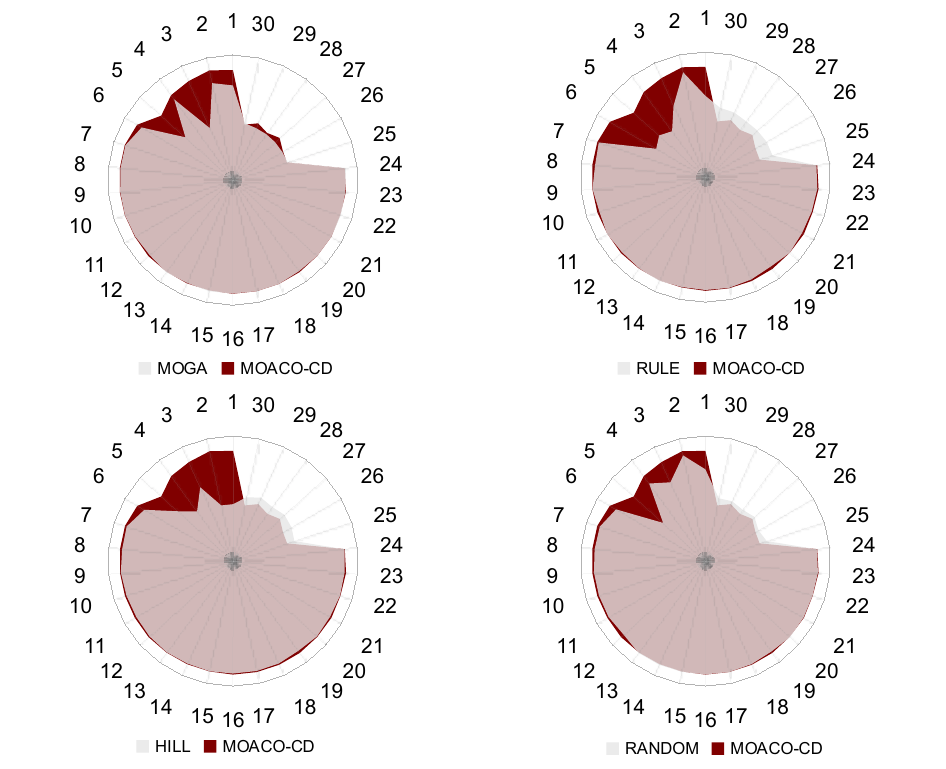}
\caption{Pair-wised comparison between MOACO-CD and each of the other approaches (the larger area means better trade-offs; objective number 1-6 are Response Time; 7-12 are Throughput; 13-18 are Reliability; 19-24 are Availability and 25-30 are Cost).}
\label{fig:trade-offs}
\end{figure}

As for G-Distance, we note that our MOACO-CD again has the best result,  producing the best quality of trade-offs in terms of the extents to which the objectives are optimized.  We can see that RANDOM is better than MOGA, RULE and HILL, this is because even though it is restricted by the weighted-sum of objectives, RANDOM tends to largely improve on a few objectives and thus leading to second best G-Distance result. The MOGA is ranked the third, because despite it caters for multi-objective, the inability to handle large number of objective and the limited diversity have caused it to optimize only a small amount of objectives. We can also see that the RULE and HILL are the worst and they exhibit marginal difference. This is because RULE is not capable to perform explicit trade-offs and optimization; while HILL is affected by overhead due to its greedy nature.

To provide a detailed view of the achieved QoS and cost values, Fig. \ref{fig:trade-offs} shows pair-wised comparisons between MOACO-CD and other approaches with respect to each of the 30 objectives that we have considered. We can clearly see that in contrast to the other autoscaling approaches, the MOACO-CD covers larger area. In particular, its decisions tend to be significantly better than those of the others on most QoS objectives while slightly worse, mainly on the Availability (against MOGA) or Cost (against RULE, HILL and RADNOM) objectives, which are smaller in number. This means MOACO-CD favors decisions that largely improve on the majority of the objectives; while causing smaller degradation to others.

In conclusion, the MOACO-CD produces better trade- offs than the others in terms of the numbers of favorable objectives and the extents to which they are optimized. This is because it favors the autoscaling decisions that do not only benefit all the objectives, but also tend to improve on each individual objective as much as possible. Therefore,  MOACO-CD is capable to perform better optimization and find trade-offs decisions with higher diversity for large number of objectives. In addition, the \emph{compromise-dominance} balances the improvements in the objectives, which leads to well-compromised trade-offs. In particular, the possible trade-offs are handled properly, not only for the naturally conflicted objectives (e.g., Throughput and cost of a service); but also for the conflicts caused by QoS interference.

\begin{table}[t!]
\centering
  \caption{The Average Violations (\%). (The Best is Highlighted in Bold)}
\label{table:violation} 
\begin{tabularx}{\columnwidth}{|P{0.7cm}cP{0.95cm}P{0.6cm}P{0.6cm}P{0.6cm}Y|}

\hline \hline
\multicolumn{2}{|c}{} &
MOACO-CD &
MOGA &
RULE &
HILL &
RAN-DOM \\ 
\hline \hline
\multirow{2}{*}{Service1} &
Response Time &
\bfseries 102.02 &
190.42 &
921.62 &
853.32 &
259.94 \\
&
Throughput &
\bfseries 10.49 &
13.45 &
14.37 &
14.52 &
14.26 \\
\hline
\multirow{2}{*}{Service2} &
Response Time &
\bfseries 86.37 &
316.74 &
2370.53 &
434.88 &
401.09 \\
&
Throughput &
\bfseries 19.15 &
20.46 &
21 &
21.86 &
21.06 \\
\hline
\multirow{2}{*}{Service3} &
Response Time &
\bfseries 99.52 &
389.12 &
405.33 &
293.55 &
457.53 \\
&
Throughput &
\bfseries 39.71 &
39.93 &
39.79 &
41.57 &
40 \\
\hline
\multirow{2}{*}{Service4} &
Response Time &
\bfseries 73.79 &
614.90 &
797.53 &
730.25 &
617.17 \\
&
Throughput &
\bfseries 19.84 &
21.33 &
20.75 &
21.58 &
19.86 \\
\hline
\multirow{2}{*}{Service5} &
Response Time &
\bfseries 0 &
186.71 &
357.58 &
676.56 &
236.49 \\
&
Throughput &
\bfseries 13.06 &
13.08 &
13.22 &
16.48 &
14.81 \\
\hline
\multirow{3}{*}{Service6} &
Response Time &
\bfseries 16.37 &
560.74 &
214.88 &
2364.84 &
192.83 \\
&
Throughput &
61.81 &
61.93 &
\bfseries 60.18 &
62.49 &
62.3 \\
&
Availability &
\bfseries 0 &
\bfseries 0 &
\bfseries 0 &
0.02 &
\bfseries 0 \\

\hline \hline 
\multicolumn{2}{|c}{Standard Deviation} &
\bfseries 0.52 &
2.69 &
6.62 &
7.55 &
2.49 \\
\hline \hline

\end{tabularx}
\end{table}

\begin{figure*}[t!]
\centering
\includegraphics[width=\textwidth]{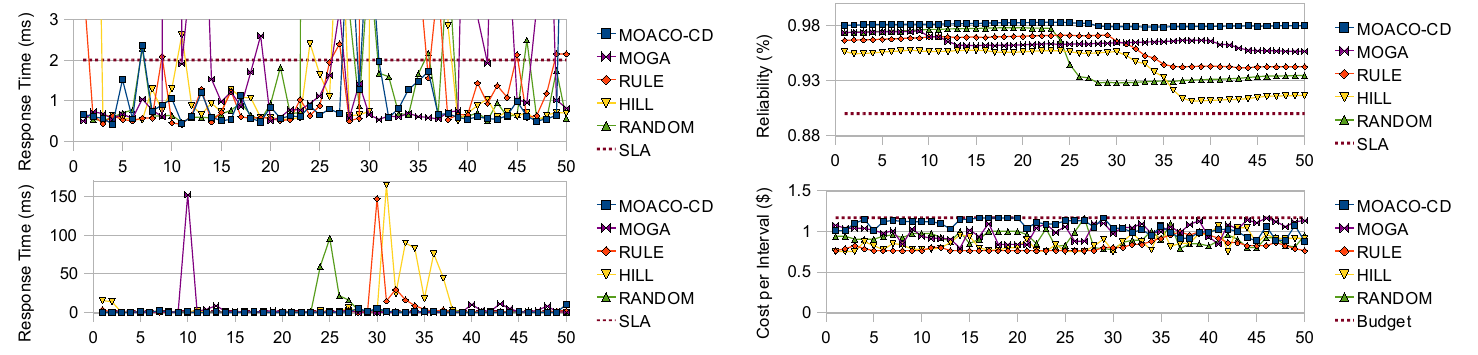}
\caption{The achieved trend of Response Time, Reliability and Cost for Service 6 in one experiment run.}
\label{fig:sla} 
\end{figure*}

\subsection{Violations of Requirements}
\label{sec:sla-evaluation}
Next, we examine whether the decisions made by our approach can eliminate runtime violations of the SLA and budget, as listed in Table \ref{table:sla-budget}. We use (\ref{eq:sla}) to assess the extents of these violations when they occur. As shown in Table \ref{table:violation}, violations do exist, mainly for the Response Time and Throughput objectives. However, we can see that MOACO-CD leads to significantly smaller violations as when compared with the others\textemdash it has the best results for 12 out of the 13 cases. In contrast, MOGA is ranked the second as it obtains the second best results for most cases. This proves that MOACO-CD outperforms MOGA in reducing SLA  violations, while optimizing and making trade-offs for large number of objectives. RANDOM is ranked the third while HILL and RULE do not differ much in terms of the overall violations. In particular, the maximum violation of MOACO-CD is only 102.02\%, which is at least 6 times better than the 2307.53\% for RULE, the 2364.84\% for HILL, the 617.17\% for RANDOM and the 614.9\% for MOGA. We can also see that MOACO-CD has the smallest standard deviation on the violations for different service-instances, meaning that the results of MOACO-CD are better balanced than the other four. This implies that the trade-offs caused by QoS interference are better compromised; otherwise, it can result in imbalanced scenarios where the QoS attributes of some service-instances are advantaged while those of the others are severely violated, e.g., as RULE and HILL.

As a detailed example, Fig. \ref{fig:sla}  illustrates the fluctuations of QoS and cost for Service 6 in one experiment run. For Response Time (Fig. \ref{fig:sla} left), we can clearly see that, in contrast to the others, our MOACO-CD does not only significantly reduce violations, but also  produce better and more stable response time when the SLA is complied. From Fig. \ref{fig:sla} (top right), we can also  observe that there are no violations for Reliability. In such case, MOACO-CD does not only constantly produces the best performance, but also tends to have the most stable results along the trend: we can clearly see that for the other approaches, the achieved reliability drops gradually at around 25-30 time step; whereas the results achieved by MOACO-CD do not fluctuate much. Finally, we can see that the cost incurred by MOACO-CD is similar to MOGA, but slightly higher than the others (Fig. \ref{fig:sla} bottom right); however, the extra cost is within the budget and it is therefore acceptable. That is to say, MOACO-CD might come with some extra costs but it can lead to significantly better results on many other QoS objectives.

In summary, we can conclude that MOACO-CD performs significantly better than the other approaches for reducing SLA violations on a large number of objectives. In addition, it leads to better and more stable results when the SLAs are complied. This might come with slightly higher cost, yet still comply with the budget requirements. Further, MOACO-CD achieves well-balanced improvement on the QoS attributes for different service-instances, which implies that the trade-offs caused by the QoS interference for both services and VMs are well compromised, even for large number of objectives.

\subsection{Elasticity}

We now evaluate elasticity of the proposed approach by means of over- and under-provisioning using (\ref{eq:provision}). Table \ref{table:provision} shows the average results for all managed service-instances and VMs on the master PM.

\begin{table}[t!]
\centering
  \caption{Over- and Under-Provisioning (\%). (The Best is Highlighted in Bold)}
\label{table:provision} 
\begin{tabularx}{\columnwidth}{|YcP{1cm}cP{0.5cm}P{0.5cm}c|}

\hline \hline
\multicolumn{2}{|c}{} &
MOACO-CD &
MOGA &
RULE &
HILL &
RANDOM \\ 
\hline \hline
\multirow{2}{*}{CPU} &
Over &
\bfseries 6.69 &
15.09 &
9.48 &
20.36 &
15.07 \\
&
Under &
 14.40 &
 9.86 &
11.82 &
13.06&
\bfseries 9.46 \\
\hline
\multirow{2}{*}{Memory} &
Over &
10.50 &
10.66 &
\bfseries 0.52 &
0.94 &
3.00 \\
&
Under &
\bfseries 2.21 &
2.50 &
19.13 &
20.40 &
11.13 \\

\hline
\multirow{2}{*}{Thread} &
Over &
22.29 &
37.05 &
\bfseries 21.94 &
32.38 &
40.00 \\
&
Under &
22.92 &
15.55 &
39.05 &
27.42 &
\bfseries 9.30 \\

\hline \hline

\end{tabularx}
\end{table}

For CPU and thread, the results of MOACO-CD do not differ much as when compared to the other four. In addition, the amounts of over-/under-provision are balanced. Interestingly, for memory, we can see that  MOACO-CD and MOGA perform significantly better than the others on under-provision, but they are the worst on over-provision with considerable difference. This is because they detect that memory can be the most critical control primitives that significantly influences the QoSs. Moreover, they have assumed that some extra costs can lead to significantly better performance on  other objectives. Consequently, both MOACO-CD and MOGA try to avoid under-provisioning by allocating more memory than the actual demand. Indeed, in contrast to MOACO-CD and MOGA, although the other three have better results on over-provision, their bigger under-provision have resulted in significantly worse QoS and SLA violations (especially for RULE and HILL), as evident in Section \ref{sec:sla-evaluation}. Finally, although MOACO-CD and MOGA obtain similar results for elasticity, we have shown that MOACO-CD outperforms MOGA on the quality of trade-offs and the ability to reduce SLA violation.

In conclusion, our approach results in good elasticity, providing that the amounts of over-/under-provision achieved by MOACO-CD are balanced and acceptable for CPU and thread. Among the others, MOACO-CD tends to have the best under-provision and the second worst, yet acceptable over-provision for memory. However, this is a trade-off between cost and QoS attributes, where the MOACO-CD has assumed that large improvements on the QoS attributes can be achieved by having slightly more costs, which are mainly spent on the memory.

\subsection{Overhead}

Finally, we  validate  the  overhead  of  our  approach  by computing latency of the decision making process. We have omitted RULE as it has negligible overheads. As shown in Table \ref{table:overhead}, we can see that for all four approaches, there is a considerable difference between the worst case and best case scenarios. Indeed, their actual overhead can be sensitive to the complexity of the used models (i.e., by the \textit{QoS Modeler}); and the number of objectives that are assigned in the same decision making process (i.e., by the  \textit{Region Controller}). For the best case scenario, our MOACO-CD has the smallest overhead (1.2s) while the HILL has the biggest. However, the results of all four approaches are acceptable. On the other hand, the RANDOM achieves the smallest overhead (38.91s) in the worst case scenario; while the MOACO-CD, MOGA and HILL report 50.3s, 69.7s and 75.09s respectively. Nevertheless, as we have seen in previous sections, the MOACO-CD is significantly better than RANDOM in terms the quality of trade-offs and its capabilities in reducing SLA violations. For both cases, MOGA has bigger latency than MOACO-CD due to the overhead of pareto-dominance sort during optimization. Another observation is that HILL is often forced to terminate as it reaches the runtime threshold (i.e., 75s); thus its actual overhead in the worst case scenario can be bigger than 75.09s. This is the main reason that causes its poor performance in the quality of trade-offs and violations. Overall, MOACO-CD has acceptable overhead even for the worst case with a sampling interval of 120s. 

\begin{table}[t!]
\centering
  \caption{The Overhead (s). (The Best is Highlighted in Bold)}
\label{table:overhead} 
\begin{tabular}{|ccccc|}

\hline \hline
&
MOACO-CD &
MOGA &
HILL &
RANDOM \\ 
\hline \hline
Best case &
\bfseries 1.2 &
12.3 &
6.8 &
3.5 \\
Worst case &
50.3 &
69.7 &
75.09 &
\bfseries 38.91 \\
\hline \hline

\end{tabular}
\end{table}

\subsection{Discussion on Complexity and Scalability}

One benefit of MOACO, as we have shown, is that it can efficiently achieve approximated results for NP-hard problems by exploring in diversified parts of the search space, leading to less effort on computing the objectives' results. 

\textbf{\textit{Lemma 1:}} Let $c_i$  be the number of possible configured values for the \emph{ith} control primitive; \emph{n} be the total number of control primitives for all the \emph{m} dependent objectives. An exhaustive search requires to perform combinatorial permutation, resulting a complexity of $O(m\cdot  \prod_{i=1}^nc_i)$. In contrast, the MOACO has a runtime complexity of $O(m\cdot  \sum_{i=1}^nc_i + m\cdot i \cdot a\cdot r)$, where \emph{i}, \emph{a} and \emph{r} are the number of iteration, the number of ants and the number of runs for an ant to find satisfactory decision, respectively. 

\textbf{\textit{Proof:}} The complexity of MOACO can be discussed in two sequential stages: updating heuristics and updating pheromone. Since the aim of updating heuristics is to calculate the quality of each individual configured value with respect to all the objectives, it is only performed once in a linear manner, resulting a complexity of $O(m\cdot  \sum_{i=1}^nc_i)$. Updating the  pheromone, on the other hand, is achieved using the decisions found by the ants and it is an iterative process throughout the algorithm. Therefore, its complexity is sensitive to \emph{m}, \emph{i}, \emph{a} and \emph{r}. This gives us the upper bounded complexity of $O(m\cdot i \cdot a\cdot r)$. In total, the MOACO has a complexity of $O(m\cdot  \sum_{i=1}^nc_i + m\cdot i \cdot a\cdot r)$. It is easy to see that, depending on the chosen  \emph{i}, \emph{a} and \emph{r}, the complexity of MOACO can be closed to or far away from that of an exhaustive search. Indeed, the optimal setting can vary according to the scenarios. However, as we have shown, our approach produces good results even with a considerably small setting of  those  parameters. 

\textbf{\textit{Lemma 2:}} Let \emph{N}, $N_p$ and $N_n$ be the number of  decisions produced by MOACO, the number of decisions filtered by pareto-dominance and the number of decisions filtered by nash-dominance, respectively. The complexity of \emph{compromise-dominance} is $O(m\cdot N^2+m \cdot N_p^2 + m \cdot N_n^2)$.

\textbf{\textit{Proof:}} Our \emph{compromise-dominance} can be discussed in three sequential sorts: pareto-dominance sort, nash-dominance sort and distance sort. We realized the pareto-dominance sort in the same way as the sort in NSGA-II, which has been proved to have a complexity of $O(m \cdot N^2)$ \cite{nsgaii}. The nash-dominance sort is similar to the  pareto-dominance sort,  with an equal or smaller number of decisions, resulting a complexity of $O(m \cdot N_p^2)$. Finally, the distance sort is based on \emph{Quick Sort} and it has a bound of $O(m \cdot N_n^2)$. Overall, we have $O(m \cdot N^2+m \cdot N_p^2 + m \cdot N_n^2)$.

The scalability of MOACO-CD is related only to the number of dependent objectives, thus it can work in large number of services because (i) increasing the number of services may not influence the approach, as our prior work \cite{Chen:2013}\cite{Chen:2014:ucc}\cite{Chen:2014} cluster the independent objectives of these services into different decision making processes, which run independently. In other words, a large number of services may not affect the decision making as long as the number of dependent objectives in a process does not change significantly. (ii) If it is known that the number of dependent objective will be largely increased, there are additional mechanisms to improve scalability, e.g., by using admission control to restrict the number of service on a VM and the VMs on a PM, which will limit the number of dependent objectives in one process. (iii) In case the  additional mechanisms are not applicable, the MOACO-CD can still be tuned using the configurations in Table \ref{table:moaco-config}. This can be achieved by profiling with respect to the possible number of dependent objectives, as we have done in this work. It is worth noting that since we consider the trade-offs caused by QoS interference, we have evaluated up to 30 objectives in one decision making process, which itself is a significantly larger scale as when compared to the small scale (i.e., 2 to 4 objectives) in existing work

\section{Related Work}
\label{sec:relatedwork}

\subsection{Rule-based Control}

In the most classic rule-based autoscaling decision making, one or more conditions are manually specified and mapped to an autoscaling decision, e.g., increase CPU and memory by \emph{x} if the throughput is lower than \emph{y}. Brandic et al. \cite{Compsac_2010_I_Brandic} propose a layered framework for autoscaling hardware resources in the cloud, in which dependency are specified between the SLA violations and the horizontal scaling decisions. Similarly, Han et al. \cite{scale-rule-based} apply multiple combination of utilizations and SLA violations to reach a decision. [19] reports on a rule-based state machine approach for autoscaling. They consider QoS interference but only focus on software control primitives. Unlike our approach, the static nature of the rules in their work requires to assume all the possible conditions/decisions and the effects of these decisions. Thus, they are limited in dealing with the dynamics and uncertainties in cloud, and resolving trade-offs.

\subsection{Control Theoretic Approaches}

Advanced control theory has also been used for autoscaling decision making in cloud because of their low latency. Among the others, \cite{acdc09} present a P-control approach to scale the number of PM for optimizing QoS in cloud based on the changes in utilization. \cite{TR-10-full-version} propose to use PI-controller for scaling CPU and memory. This controller calculates "error" values as the difference between hardware control primitives and throughput. \cite{qcloud} present another classical controller that is specifically designed to handle VM-level QoS interference in cloud. More  sophisticated controller also exist, for instance, Fuzzilized control \cite{IWQOS11} has been studied for autoscaling hardware resources in cloud. However, the major drawback of control theoretic approaches is that they require to make many actuations on the physical system, in order to collect the "error" for stabilizing itself. In contrast, our approach uses search-based optimization via QoS models, and thus it permits to do extensive reasoning before the actual actuations are taken placed. In addition, control theory is difficult to be adapted for effectively reasoning and resolving trade-offs at runtime.

\subsection{Search-based Optimization}

The proposed approach also belongs to the broad category of search-based optimization, in which the decisions are extensively reasoned in a finite, and possibly large search space. Search-based optimization for autoscaling decision making in the cloud can be either \emph{explicit} or \emph{implicit}. The former performs optimization as guided by explicit system models; while this process is not required for the later. As examples of the implicit approaches, \cite{parallel-RL-vertical-QoS-2013} and \cite{2013-JRAO-most-closest-work-2013} use model-free reinforcement learning algorithm to make autoscaling decision with respect to the hardware control primitives. In particular, \cite{2013-JRAO-most-closest-work-2013} have considered software control primitives and the QoS interference. Approaches that rely on demand prediction are also belong to this category, e.g., the CloudScale \cite{cloudscale}. Nevertheless, the implicit approaches cannot explicitly handle the trade-offs. 

As for the explicit ones, \cite{multitier-resalloc-Cloud11}\cite{2014-2-SVM-workload-type-2014}\cite{ILP-cost-only-scaling-2013} attempt to scale control primitives in the cloud with a single objective in mind (e.g., the cost). The optimization algorithms include, e.g., force-directed optimization \cite{multitier-resalloc-Cloud11}, lagrange algorithm \cite{2014-2-SVM-workload-type-2014} and integer optimization \cite{ILP-cost-only-scaling-2013}. However, they tend to be limited in making trade-offs, because the search is restricted to favor a single objective only. Many other researches have been conducted using a weighted-sum of objectives. The optimization algorithms range from simple techniques, e.g., Exhaustive Search \cite{cache-static-ANN-bi-obj-2012} and Decision Tree Search \cite{2014-profiling-decision-tree-scaling-2014}, to complex ones, e.g., Hill-climbing Search \cite{smart2004} and Genetic Algorithm  \cite{software-RP-two-loops}\cite{Antonescu13dynamicsla}.  In particular, \cite{ant-cloud} propose to use Ant Colony for autoscaling hardware resources based on weighted-sum of objectives.  Nevertheless, weighted-sum of objectives tends to restrict the search and has limited diversity in the produced decisions. In addition, they do not consider the trade-offs caused by QoS interference.

There is a limited amount of work that uses multi-objective optimization in cloud autoscaling and they have assumed no more than 4 objectives. \cite{GA-full-simulation} have used MOGA that derived from NSGA-II for trade-off decisions in autoscaling. Similarly, \cite{E3-R-extended} have also leveraged on NSGA-II but they have additionally considered the reduction of harmonic QoS objectives. \cite{2014-eplison-GA-weigh-h-scaling-2014} is another example that use NSGA-II, but the authors apply a weaker form of pareto-dominance, namely epsilon dominance, where it requires a static epsilon value as the minimum number of better objectives that a non-dominated decision needs to achieve. However, all these approaches apply pure pareto-dominance to evaluate the overall quality of decisions for all the objectives during the optimization, which will lead to a large set of  non-pareto-dominated decisions. Such fact can limit NSGA-II when the number of objectives increases \cite{yao}. Moreover, they cannot guarantee well-compromised trade-offs and thus can lead to imbalanced trade-offs. They have also ignored software control primitives and the trade-offs that caused by QoS interference.  In contrast, we have explicitly considered QoS interference, and by using aggregative heuristics and different pheromone structures for the objectives, we design MOACO in a way that similar to conduct many single objective optimizations in one run. Therefore, instead of evaluating the overall quality of decision for all objective during the optimization (as in MOGA), each ant  assess the decision against single objective, and this avoids the use of pareto-dominance in optimization. The overall quality of decisions are evaluated using \emph{compromise-dominance} upon completion of the optimization, and this ensure well-compromised trade-offs. Such design aims to optimize and to make trade-offs for large number of objectives while ensuring good diversity. \cite{ant-cloud-vm} is a related work to our approach as they also applied MOACO in cloud. However, their focus is on the trade-offs  for VM to PM mapping problem, whereas ours is on the trade-off decision making problem in autoscaling. In addition, they have assumed only two objectives; while our approach does not assume such limit and we have used up to 30 objectives in the experiments. They update pheromone by using pareto-dominance to evaluate decisions against all objectives during the optimization; whereas we conduct the update by assessing the decisions for each objective. After the optimization completes, we explicitly search for well-compromised trade-offs using \emph{compromise-dominance}.

\section{Conclusion and Future Work}
\label{sec:conclusion}
In this paper, we present a self-adaptive approach for autoscaling decision making in the cloud. In particular, it adaptively resolves the trade-offs without human intervention.  By leveraging on MOACO, the approach dynamically searches and optimizes for possible trade-offs with high diversity. Further, we propose compromise-dominance for adaptively selecting the decision that leads to well-compromised trade-offs. The experiments show that,  in contrast to the rule-based, heuristic based, randomized and MOGA based autoscaling approaches, our approach produces better trade-offs quality in terms of the numbers of favorable objectives and the the extents to which they are optimized; and much smaller violations of the requirements with large number of objectives. Moreover, it results in acceptable overhead  and has balanced elasticity in terms of the over-/under-provision. 

In future work, we plan to investigate how the latency of scaling actions (e.g., replication and migration) can affect the decision making process. Extending the work for managing energy in cloud is also in our ongoing agenda.

% or


%\appendices
%% you can choose not to have a title for an appendix
%% if you want by leaving the argument blank
%\section{}
%% use section* for acknowledgment
%\ifCLASSOPTIONcompsoc
%  % The Computer Society usually uses the plural form
%\else
%  % regular IEEE prefers the singular form
%\fi

\ifCLASSOPTIONcaptionsoff
  \newpage
\fi



\bibliographystyle{IEEEtran}
 %argument is your BibTeX string definitions and bibliography database(s)
\bibliography{references}
%\begin{thebibliography}{1}
%\end{thebibliography}

\vspace*{-2.6\baselineskip}
\begin{IEEEbiography}[{\includegraphics[width=1in,height=1.25in,clip]{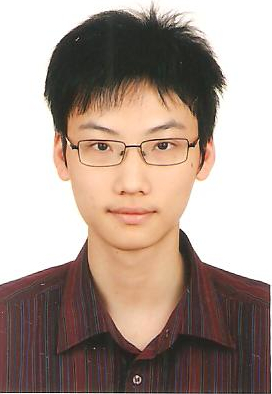}}]{Tao Chen}
is a Research Fellow at the School of Computer Science, University of Birmingham, UK. His research interests include performance/QoS modeling and tuning, self-adaptive systems, software engineering, cloud computing, services computing and distributed computing. His work has been published in \emph{SEAMS}, \emph{UCC}, \emph{INS}, \emph{Computer}, \emph{IEEE Cloud} and \emph{ICCS}.
\end{IEEEbiography}
\vspace*{-2\baselineskip}
\begin{IEEEbiography}[{\includegraphics[width=1in,height=1.25in,clip]{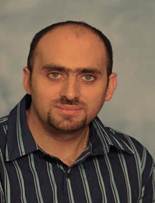}}]{Rami Bahsoon}
is a Senior Lecturer in Software Engineering and founder of the \textit{Software Engineering for/in the Cloud} interest groups at the School of Computer Science, University of Birmingham, UK, working in areas related to cloud software engineering and economics-driven software engineering and architecture. He is currently acting as the workshop chair for \emph{IEEE Services} 2014, the Doctoral Symposium chair of \emph{IEEE/ACM UCC} Conference 2014, and chair for the visionary track of \emph{IEEE Services} 2015.
\end{IEEEbiography}
%\vspace*{-1\baselineskip}
%% insert where needed to balance the two columns on the last page with
%%\newpage


%\vfill


\end{document}